\documentclass[usenatbib]{mn2e}
\usepackage{graphicx}

\usepackage{amssymb}
\usepackage{color}
\usepackage{float}
\usepackage{amsmath}
\usepackage[title]{appendix}
\newcommand{\bn}{\begin{enumerate}}
\newcommand{\en}{\end{enumerate}}
\newcommand{\bi}{\begin{itemize}}
\newcommand{\ei}{\end{itemize}}

\usepackage{xcolor}

\def\gtorder{\mathrel{\raise.3ex\hbox{$>$}\mkern-14mu
    \lower0.6ex\hbox{$\sim$}}}
\def\ltorder{\mathrel{\raise.3ex\hbox{$<$}\mkern-14mu
    \lower0.6ex\hbox{$\sim$}}}

\newcommand{\apj}{ApJ}
\newcommand{\aap}{A\&A}
\newcommand{\apjl}{ApJL}

\newcommand{\mnras}{MNRAS}

\newcommand{\nat}{Nature}

\title[Violent Buckling Benefits Galactic Bars]{Violent Buckling Benefits Galactic Bars} 
\author[Angela Collier]
{Angela Collier$^{1}$\thanks{E-mail: angela.collier@colorado.edu}
\\
$^{1}$ JILA and Department of Astrophysical and Planetary Sciences, CU Boulder, Boulder, CO 80309, USA\\
}

\begin{document}

\date{Accepted ?; Received ??; in original form ???}

%\pagerange{\pageref{firstpage}--\pageref{lastpage}} \pubyear{2014}

\maketitle

\begin{abstract}

Galactic bars are unstable to a vertical buckling instability which heats the disk and in some cases forms a boxy/peanut shaped bulge. We analyze the buckling instability as an application of classical Euler buckling followed by nonlinear gravitational Landau damping in the collisionless system. We find that the buckling instability is dictated by the kinematic properties and geometry of the bar.  The analytical result is compared to simulations of isolated galaxies containing the disk and dark matter components. Our results demonstrate that violent buckling does not destroy bars while a less energetic buckling can dissolve the bar. The disks that undergo gentle buckling remain stable to bar formation which may explain the observed bar fraction in the local universe. Our results align with the results from recent surveys.
\end{abstract}

%%%%%%%%%%%%%%%%% END OF PREAMBLE %%%%%%%%%%%%%%%%


%%%%%%%%%%%%%%%%%%%%%%%%%
\section{Introduction}
\label{sec:intro}

Over $70\%$ of disk galaxies host galactic bars. For a review of observations and bar fraction measurements, see \citet{erw18} and the references therein. The galactic bar is important for galaxy morphology and evolution. It not only shapes the disk; the bar is responsible for angular momentum transfer throughout the galaxy. The earliest simulations of isolated systems show bars forming spontaneously in axisymmetric disks \citep[e.g.,][]{ost73,hohl}.  
Simulated bars can also form from tidal interactions \citep[e.g.,][]{nog87,ger}.  
The ease with which simulated disks form bars and their pervasiveness in observations raises the question, how can some disks avoid bar formation?

Regardless of the origin of the bar, the increase in radial dispersion velocities along the growing bar makes the disk increasingly unstable to buckling modes. The bar quickly undergoes a second instability which thickens and heats the disk vertically. The buckling instability first appeared in simulations when three dimensional models were evolved \citep[e.g.,][]{combes90,pfenn91,raha91}. Following the discovery of buckling in thin disks, the structure of the orbit families of the three dimensional bar were also studied \citep[e.g.,][]{pfenn91,skokos02,marti04}. Bar orbits develop vertical structure (such as banana and anti-banana orbits)  to support the thickened bar. 

Edge-on the bars sometimes have a 'boxy/peanut' shaped bulge (B/P bulge) often attributed to the buckling instability \cite[e.g.,][]{pfenn90,raha91,atha02,marti04,deba06,saha13}. 
\citet{atha02} found that the B/P appearance continues to grow in size along with the bar, long after buckling has subsided. The Milky Way bar is believed to have buckled in the past and formed the observed 'X-shaped' bulge \citep[]{shen10,marti11}. Surveys have found the buckling could plausibly account for the large fraction of B/P bulges in the local universe \citep{erw16}.

Therefore, buckling is of great importance to galactic evolution and deserves further study. Recent works have begun to highlight how halo angular momentum can contribute to stellar disk evolution, particularly its effect on the long term evolution of bars. While the angular momentum found in dark matter halos is inconsequential dynamically \citep{bull01}, there are many effects seen in bar evolution \citep[e.g.,][]{sahanaab13, coll1, coll2, coll3}. 
First, the more rotation in the halo the shorter the timescale of the initial bar instability. Secondly and more related to this work, the buckling instability leads to a larger reduction in bar strength when halos have angular momentum. Finally, in halos of high spin the bar is dissolved during the buckling and does not reform within the disk for the rest of its evolution. \citet{coll2} studied the long term evolution of the resulting bar dissolution in great detail and found it to be due to the formation of a relatively strong dark matter or 'ghost' bar found in spinning halos. The trapping of dark matter material by the stellar bar changes the mass, shape and orbital structure of the galactic bar which in turn changes the initial conditions of the buckling instability.

The discussion of buckling is usually limited to the resulting bar structure as there is no complete analytical description of buckling instability that predicts the time or mechanics of its onset within a barred galaxy. Additionally, the timescale of the instability is quite short compared to bar lifetimes.   Here we discuss the detailed physical processes involved in the buckling instability by studying what bar properties effect the initial buckle as well as how the perturbation moves through the system.  We provide a description of the mixing that washes out the buckling perturbation. We ask; what bar properties predict the initial buckling deflection from the plane? How does the shape of the buckling perturbation effect the time scale and energetics of the event? And finally, how do different buckling profiles effect the long term evolution of the disk? 

It must be mentioned that \citet{marti06} have shown that a bar that grows substantially in length may undergo a second buckling instability which has structurally different evolution than the first buckling. The second buckling instability will be discussed in a future work.  

This paper is structured as follows. The two phases of the buckling instability are described in Section \ref{sec:buck}. Next, simulations are outlined in Section \ref{sec:ICs} and the results are presented in Section \ref{sec:results}. We discuss the simulations in the context of the analytical results in Section \ref{sec:discussion} followed by a brief summary of conclusions and comparison to recent observations in Section \ref{sec:conclusion}. 

 \begin{figure}
\centerline{
 \includegraphics[width=0.5\textwidth,angle=0] {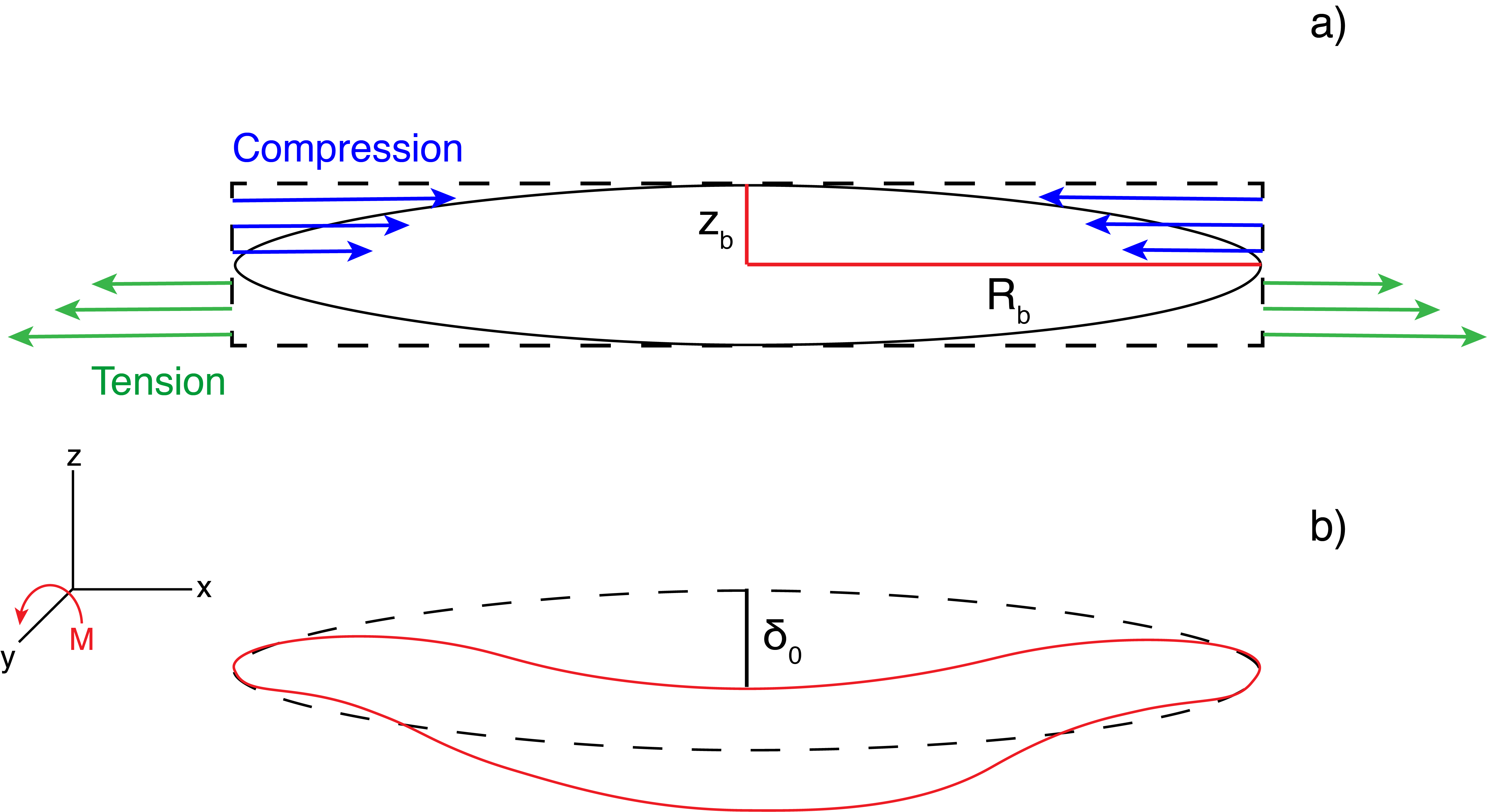}}
\caption{a. Indicates the direction of compression and tension on the bar just before buckling. These forces cause the bar to bend and the magnitude of these forces determine the buckling moment of force ($M$). 2. Cartoon of buckled bar. Direction of buckling moment of force is shown. Note at the instantaneous time of buckling the bars edges remain pinned to the galactic plane conserving $R_b$. The maximum deflection is marked as $\delta_o$.}
\label{fig:buck_cart}
\end{figure}

\section{The Buckling Instability}
\label{sec:buck}
After the bar instability breaks the symmetry of the disk the bar grows in length and strength by trapping additional stellar orbits. The initial edge-on buckling perturbation in the disk develops spontaneously. \citet{toomre66} analyzed buckling modes in idealized, thin sheets with uniform density and found the sheets stable if the ratio of vertical dispersion velocities ($\sigma_z$) to radial dispersion velocities  ($\sigma_r$) is $\gtorder{0.3}$. Outside this range the buckling modes appear in the thin sheets and are compared to the fire-hose instability of plasma physics.  This idealized analysis does not apply strictly to bars but it is appropriate to follow the radial and vertical dispersion velocities in the bar due to the drastic increase in vertical dispersion velocity during buckling. In isolated systems, the bar forms in the initially thin disk which increases the radial dispersion velocities while the vertical dispersion velocities remain unchanged until the eventual buckling. This has led to other works discussing the importance of $\sigma_r$ alone in kinematic fractiation of the edge on disk \citep{deba17}.

An alternate instability was suggested by \citet{combes90} and \citet{pfenn91} after studying the three-dimensional orbits of stars in barred potentials. They found that orbits close to a 2:1 vertical resonance are unstable causing the bar to bend out of the plane. \citet{merr94} however, stressed that this type of instability cannot fully explain buckling. While not all stars lie close to the resonance, all stars trapped in the bar are perturbed by the initial buckling mode.

Once the buckling perturbation appears the bar oscillates above and below the plane until the buckling is washed out. Both phases of the buckling instability are described below. The second phase of buckling will be analyzed as a  form of Landau damping \citep{land46}.  Landau damping as a method of collisionless decay of oscillations was initially prescripted for plasmas but can be converted to describe gravitational systems as well (e.g., \citet{lynd62} \citet{m91} and \citet{k98}). Gravitational Landau damping will arise when individual particles interact with a wave potential (such as a buckling wave). With Landau damping, particles with a velocity larger than the wave will lose energy to the wave while particles with a lower velocity will gain energy which damps the wave. \citet{binn08} make the analogy of a surfer riding an ocean wave. When the surfer passes behind the crest of the wave they do work on the wave but when the surfer points the board in the downhill direction of the wave the wave does work on the surfer. To extend this analogy, the closer the surfer's velocity is to the velocity of the wave the longer the surfer can hold the energy gained by the wave. This is true for Landau damped particles as well. The Maxwellian distribution of particle velocities implies that some particles will have a velocity near the velocity of the perturbation and some will be very far away from it. Those far away will gain and lose energy quickly while those particles with energy near the perturbation will 'hold' the energy given to them by the wave for a longer time. Therefore, a perturbation with a velocity near the mean of the system velocity will take longer to damp than a perturbation with a much larger velocity.

We separate the buckling instability into two phases, an instantaneous breaking of symmetry in the x-z plane followed by the oscillation and subsequent washing out of the density perturbation.

\subsection{Symmetry breaking by buckling}
\label{sec_buck}
The galactic bar feels increasing internal pressures until it self-buckles under self-gravity. Just before the buckling instability the bar is aligned with the plane of the thin disk so the initial buckling perturbation is that of a bar pinned at both ends with internal pressure forces acting axially. Figure \ref{fig:buck_cart} depicts the coordinate system and position of the compression and tension forces on the bar at the moment of buckling. The bottom of Figure \ref{fig:buck_cart} exaggerates the bending perturbation of the bar as it buckles. In order to continue to to grow in strength and length the bar must thicken via the buckling instability to support itself.

The axial stress force on the bar that eventually leads to buckling can be approximated from dispersion velocities.  The stress tensor of a stellar fluid ($\tau$) is related to the dispersion velocities by \citet{binn08} where
\begin{eqnarray}
\tau=-\rho(r)\sigma_{i,j}^2.
\end{eqnarray}

The volume density is $\rho(r)$ and $\sigma_{i,j,}$ is defined as
\begin{eqnarray}
\sigma_{i,j}^2=\langle (v_i-\langle v_i \rangle)(v_j - \langle v_j \rangle)\rangle 
\end{eqnarray}
 
 where $i,j=r,\phi,z$ are in cylindrical coordinates. We separate $\tau$ into the normal and shear stresses 
 \begin{eqnarray}
 \label{stress}
\tau=\tau_n+\tau_s=-\rho(r)(\sigma_{ii}^2+\sigma_{ij}^2).
\end{eqnarray}
 
 It follows that for the galactic bar, the ratio of stress to strain is $\tau_s/\tau_n$. We compare the buckling of the galactic bar to the classical Euler buckling of columns where the critical stress that buckles the column is proportional to Young's modulus, the ratio of stress to strain in the material. We suggest the energy of the buckling instability should depend on the ratio of internal stresses felt by the bar at the time of buckling. 
 
 The geometric properties of the bar should play a role in the buckling as well. In the classical Euler buckling instability of columns the force of buckling is inversely related to the square of the slenderness of the column. The slenderness ratio of the bar ($R_b/z_b$) is also important in galactic buckling but we predict that a more slender galactic bar will have a more dramatic/stronger buckling when compared to a thicker bar. This prediction is based on a similar orbital inclination instability found in thin disks. \citet{mad18} found that the disk inclination instability is driven by torques between orbits which are stronger when the disks are thinner. Comparing this to a stellar bar, we expect the orbits within a thicker bar to experience less torques resulting in an overall weaker buckling. The energy of the buckling is reflected in the deflection from the plane, the more slender bar will have a larger deflection from the plane (due to the increased torques increasing the energy of the buckling) when compared to a thicker bar. The maximum deflection of the bar, $\delta_o$, is labeled in Figure \ref{fig:buck_cart}. This parameter should increase with buckling energy as energy increases with the square of the amplitude of the sinusoidal buckling force. We predict the kinematic and geometric properties of the galactic bar have an effect on the initial buckling wave with the energy of buckling increasing with internal stresses and bar slenderness.

\subsection{Landau damping of the buckling wave}
The buckling causes a sinusoidal force to be exerted onto the stars and then the buckling perturbation evolves as a wave oscillating above and below the plane. To analyze the damping of the system we describe each star as a harmonic oscillator perturbed by this force of amplitude $\delta$ and frequency $\Omega$. Both of these properties will vary with initial conditions of the buckling wave described in Section \ref{sec_buck}.  This analysis is done in the rotating  frame and we ignore motion of the stars in the radial direction. The equation of motion of a single oscillator in the presence of the buckling oscillation is

\begin{eqnarray}
\begin{aligned}
\ddot{z}+\omega^2z=\delta_o \textrm{cos}(\Omega t)
\end{aligned}
\end{eqnarray}

where $\omega$ is the vertical oscillation frequency of the star above and below the plane. Every harmonic oscillator has a unique $\omega$ due to the continuous distribution of velocities in the $z$ direction. The general solution to the equation of motion is

\begin{eqnarray}
\begin{aligned}
z(t)= z_o\textrm{cos}(\omega t)+\dot{z_o}\frac{\textrm{sin}(\omega t)}{\omega}+\\ \frac{\delta_o}{\omega^2-\Omega^2}[\textrm{cos}(\Omega t)-\textrm{cos}(\omega t)].
\end{aligned}
\end{eqnarray}

To study the response of the stars to the buckling we ignore the first two terms without $\Omega$. These terms relate to decoherence among the stellar ensemble while we focus solution to the response ro the perturbation ($z_{res}$):

\begin{eqnarray}
z_{res}(t)= \frac{\delta_o}{\omega^2-\Omega^2}[\textrm{cos}(\Omega t)-\textrm{cos}(\omega t)].
\end{eqnarray}

We calculate the alignment along the plane for the entire bar by averaging over all stars

\begin{eqnarray}
\label{eqn:222}
\langle z_{res}(t) \rangle=\delta(t)= \delta_o \int_{-\infty}^{\infty} d\omega \frac{\rho(\omega)}{\omega^2-\Omega^2}[\textrm{cos}(\Omega t)-\textrm{cos}(\omega t)].
\end{eqnarray}

As the perturbation gets damped $\delta(t)$ goes to zero. The distribution of frequencies ($\rho(\omega)$) is centered around some $\bar{\omega}$ and the frequency of the perturbation must be on the order of this average oscillation.  Consequently, we can expand Equation \ref{eqn:222} with $\omega=\Omega+(\omega-\Omega)$ and approximate the average displacement from the plane as

\begin{eqnarray}
\begin{aligned}
\label{eq:2222}
\delta(t) \approx \frac{\delta_o}{2\bar{\omega}}\left[ \textrm{cos}(\Omega t)\int_{-\infty}^{\infty} d\omega \rho(\omega)\frac{1-\textrm{cos}(\omega-\Omega)t}{\omega-\Omega}\right]\\
+\frac{\delta_o}{2\bar{\omega}} \left[\textrm{sin}(\Omega t) \int_{-\infty}^{\infty} d\omega \rho(\omega)\frac{\textrm{sin}(\omega-\Omega)t}{\omega-\Omega} \right].
\end{aligned}
\end{eqnarray}

We assume that the distribution around $\bar{\omega}$ is small which reduces Equation \ref{eq:2222} to 

\begin{eqnarray}
\begin{aligned}
\delta(t) \approx \frac{\delta_o \textrm{sin} (\bar{\omega}t)}{\bar{\omega}}\left [\int_{-\infty}^{\infty} d\omega \rho(\omega)\frac{\textrm{sin} \frac{1}{2}(\omega-\Omega)t}{\omega-\Omega}\right].
\end{aligned}
\end{eqnarray}

A single oscillator with frequency $\omega$ will find itself vibrating due the buckling perturbation according to

\begin{eqnarray}
\begin{aligned}
\label{eq:time}
z(t) \approx \frac{\delta_o\ \textrm{sin} \frac{1}{2}(\omega-\Omega)t}{\bar{\omega}\ (\omega-\Omega)}.
\end{aligned}
\end{eqnarray}

Hence, a star with frequency $\omega$ will be excited after the initial buckling and will move to a position $z=\delta_0/[\bar{\omega} (\omega-\Omega)]$ on a timescale of $t=\pi/[\omega-\Omega]$ and then will align with the plane after $t\approx2\pi/[\omega-\Omega]$. This star will experience repeat oscillations of lower amplitude as the perturbation gets damped. For stars with $\omega$ very near or equal to $\Omega$ the energy given to the star by the wave is held for a longer time compared to stars with $\omega$ very far from $\Omega$ which is a fundamental property of gravitational Landau damping. \citep{lynd62} Initially, nearly all stars in the bar contribute to the oscillation but this fraction gets lower as $t$ goes to infinity. Eventually, only stars with $\omega=\Omega$ contribute to the oscillation---if those stars exist.

At the time of the initial perturbation all oscillators are driven in phase with the initial displacement. As the instability progresses more oscillators become out phase and no longer interact with the perturbation which causes destructive interference. This leads to the perturbation decaying rapidly. The energy of the perturbation is converted into incoherent motion associated with the oscillations about the plane. The timescale of the Landau damping will be effected by the bar parameters at the time of buckling which determine $\delta_o$ and $\Omega$. A higher energy wave will have a larger amplitude and frequency and with each oscillation a larger fraction of energy will be moved from the wave motion to the individual particle motion. The opposite is true for lower energy waves. If the initial $\Omega$ of the perturbation is lower and closer to the $\bar{\omega}$ of the system the change in energy per oscillation will be smaller. In this case, the individual oscillators will spend more time in phase with the perturbation which lengthens the damping time scale. The time scale is also lengthened by the lower frequency as the time to complete one oscillation increases.

\begin{figure}
\centerline{
\includegraphics[width=0.5\textwidth,angle=0] {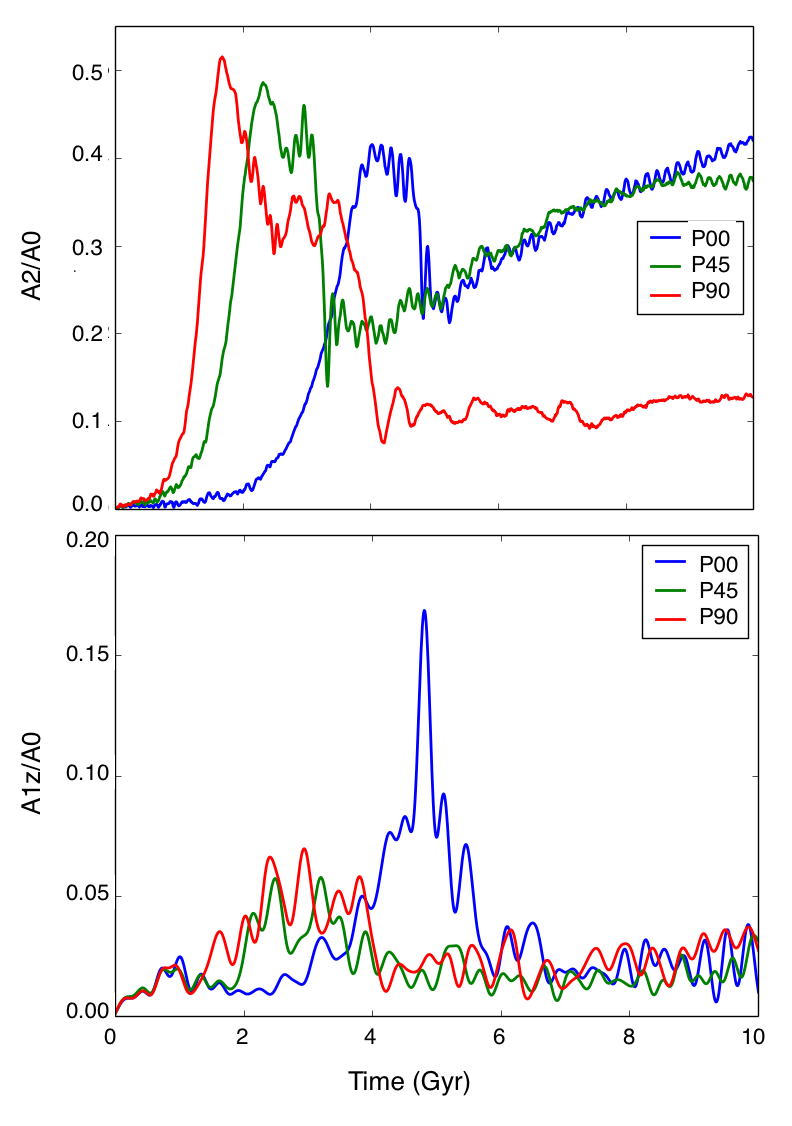}}
\caption{Top: Evolution of the bar strength parameter ($A_2/A_0$) in the x-y plane which compares the growth of the Fourier $m=2$ mode to $m=0$ mode the for three analyzed models. Bottom: Evolution of the ratio of the Fourier $m=1$ mode to the $m=0$ mode ($A_{1z}/A_0$) in the x-z plane for three galaxy models. This quantity represents the growth in asymmetry of the edge-on bar.}
\label{fig:aa}
\end{figure}

\section{Numerics}
\label{sec:ICs}
The issue of bar buckling in the context of N-body simulations of disk galaxies is now addressed.  We model stellar disks inside a spherical Navarro-Frenk-White halos \citep[hereafter NFW]{nava96} using the $N$-body part of the tree-particle-mesh Smoothed Particle Hydrodynamics (SPH/$N$-body) code GIZMO \citep{hop15}.  Our code units for mass, distance, and time are  $10^{10}\,M_\odot$, 1\,kpc, and 1\, Gyr. The time step between snapshots is shortened to $0.001$\, Gyr during the buckling in order to better study the instability which happens on quite a short timescale.

The simulations presented in this work are adapted from the models described in \citet{coll1,coll2,coll3}. We refer the reader to these works for more analysis on the long term evolution of these models.

\subsection{Initial Conditions of Galaxy Models}

We have created a series of two-component isolated galaxies with identical disks inside halos of different rotation. The halo density follows the NFW profile,

\begin{equation}
\rho_{\rm h}(r) = \frac{\rho_{\rm s}\,e^{-(r/r_{\rm t})^2}}{[(r+r_{\rm c})/r_{\rm s}](1+r/r_{\rm s})^2},
\end{equation}
where $\rho(r)$ is the DM density in spherical coordinates, $\rho_{\rm s}$ is the fitting density parameter, and $r_{\rm s}=9$\,kpc is the characteristic radius, where the power law slope is $-2$, and $r_{\rm c}$ is a central density core where $r_{\rm c}=1.4$\,kpc. The Gaussian cutoff is applied at $r_{\rm t}=86$\,kpc for the halo. The DM halo contains $7.2\times 10^6$ particles and the halo mass is $M_{\rm h} = 6.3\times 10^{11}\,M_\odot$.

The halo velocities are found by using a version of the iterative method from \citet{rodio06}, see also \citet{rodio09}. The iterations create a halo with an isotropic velocity distribution while cosmological halos are found to have a range of spin which can be fit by a lognormal distribution,

\begin{equation}
P(\lambda) = \frac{1}{\lambda (2\pi\sigma_{\lambda})^{1/2}} \textrm{exp} \bigg[-\frac{\textrm{ln}^2 (\lambda/\lambda_0)}{2\sigma_{\lambda}^2}\bigg],
\end{equation}
where $\lambda_0=0.035\pm 0.005$ and $\sigma_{\lambda}=0.5\pm 0.03$ are the fitting parameters \citep[][]{bull01}. 

To add angular momentum to the halo a fraction of randomly chosen halo particles have their tangential velocities reversed. The new velocity distribution maintains the solution to the Boltzmann equation and does not alter the velocity profile \citep{lynd60,wein85,coll1,coll2,coll3}, so the equilibrium state is preserved.

While our halos vary in spin, each simulation begins with an identical disk. The volume density of the exponential stellar disk is;

\begin{eqnarray}
\rho_{\rm d}(R,z) = \bigl(\frac{M_{\rm d}}{4\pi h^2 z_0}\bigr)\,{\rm exp}(-R/h) 
     \,{\rm sech}^2\bigl(\frac{z}{z_0}\bigr),
\end{eqnarray}
where $M_{\rm d}$ is the disk mass, $h=2.85$\,kpc is its radial scalelength, and $z_0=0.6$\,kpc is the scaleheight.  The stellar disk has $0.8\times 10^6$ particles and the disk mass is $M_{\rm d} = 6.3\times 10^{10}\,M_\odot$. The initial disk velocities depend on the potential of both components of the galaxy and are assigned last. The radial and vertical dispersion velocities are assigned as exponentials.

The galaxy simulations presented here vary only in halo spin. Following the notation of \citet{coll1,coll2,coll3} the models are labeled as $P$ for prograde rotation and then multiplied by $\lambda$ of the halo. For example, the model P00, is the nonspinning halo with $\lambda=0$. 

 \begin{figure}
\centerline{
 \includegraphics[width=0.5\textwidth,angle=0] {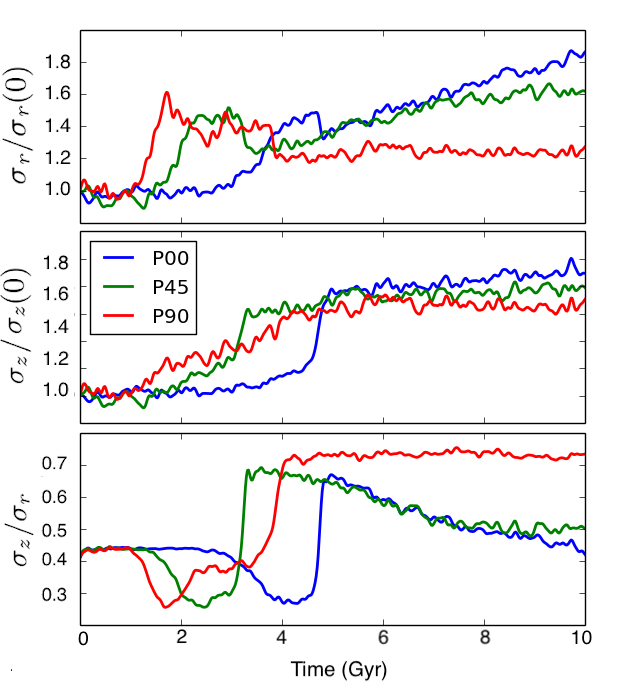}}
\caption{Top: Evolution of the radial dispersion velocities normalized by $\sigma_r(t=0)$, Middle: Evolution of vertical dispersion velocities normalized by $\sigma_z(t=0)$, Bottom: Evolution of the ratio of vertical to radial dispersion velocities ($\sigma_z/\sigma_r$). For all plots the data was measured for particles at the radii between $3-6$ kpc.}
\label{fig:disp}
\end{figure}

\section{Results from Numerical Models}
\label{sec:results}
The following analysis is done considering the trapped dark matter component as part of the total galactic bar. The dark matter component contributes to the mass, strength, and height of the bar. (e.g. \citet{coll2, coll3})  

All three models produce strong bars before buckling. The strength of the bar is defined by the ratio of the Fourier $m=2$ mode to the $m=0$ mode,

\begin{eqnarray}
\frac{A_2}{A_0} = \frac{1}{A_0}\sum_{i=1}^{N_{\rm d}} m_{\rm i}\,e^{2i\phi_{\rm i}},
\end{eqnarray} 
which is achieved by summing over all bar particles with $R \leq 14$ kpc, and mass m = $m_i$ at azimuthal angle $\phi_i$. 

We plot the bar strength for all three models in the top of Figure \ref{fig:aa} for the entirety of the simulation. While this paper strictly focuses on buckling event itself, note that the long term evolution of these models varies after buckling. This effect is discussed in \citet{coll1, coll2} in great detail. Briefly, these models vary only in $\lambda$ and the disks are identical at $t=0$. The resulting changes in bar strength evolution can be seen in the top of Figure \ref{fig:aa}. The disks start the simulations axisymmetric. When the disk undergoes the bar instability there is an exponential growth in bar strength that includes the trapping of stellar and halo particles in the bar. The bar reaches some peak strength and then undergoes the buckling instability which reduces the strength of the bar and in some models dissolves the bar, i.e. P90. In halos of lower spin the bar recovers after buckling and continues to grow in strength for the remainder of the simulation.

Halo spin effects the evolution of the bar in the following ways. First, the timescale of the bar instability is shortened in spinning halos. For example, the bar in the P90 model appears nearly two Gyrs before the bar in the P00 model. Secondly, change in strength during the buckling in stability ($\Delta A_2/A_0$) increases with increasing halo spin. The P90 bar loses $\sim 80 \%$ of its strength after buckling while the P00 bar loses only $\sim 40\%$. Third, we see the time scale of the buckling instability increases with increasing halo spin. By buckling time scale ($t_b$) we are referring to the time between the maximum value of $A_2/A_0$ just before the minimum of $A_2/A_0$. For the P90 model, $t_b=0.72$ Gyrs and for the P00 model, $t_b=0.2$ Gyrs. Finally, the post buckling evolution depends greatly on the spin of the halo, with the bar dissolving in P90. Obviously, this effect must be intensely intertwined with the dynamics of the buckling instability.

The growth of the Fourier $m=1$ mode compared to the $m=0$ mode represents the growth in asymmetry in the x-z plane. As the bar buckles out of the plane we expect to see this parameter increase in strength. 
 We plot this for each model in the bottom of Figure \ref{fig:aa}. The peak in this parameter generally matches the time of the peak in $A_2/A_0$ just before the buckling instability.

As discussed in the Section \ref{sec:intro}, the dispersion velocities in the radial direction increase as more and more stars become trapped in the bar. In the $z$-direction the orbits have random motions and oscillate above and below the plane. While the bar is growing in the disk and increasing the radial dispersion velocities; the vertical dispersion velocities do not increase until the onset of the buckling instability in these isolated models. Figure \ref{fig:disp} shows the time evolution of the radial and vertical dispersion velocities normalized by their $t=0$ values, as well as the ratio between these two measures.  This plot was taken for all time steps and includes all stellar and DM particles in the bar at the radii between $|3-6|$ kpc to avoid anything odd inside the very central bar.

In the top of Figure \ref{fig:disp} we plot the evolution of $\sigma_r$ normalized by $\sigma_r(t=0)$ for the three models. The growth of $\sigma_r$ is closely tied to the bar evolution as trapping stars in the bar increases the fraction of radial orbits. The evolution of $\sigma_r$ appears to mirror the evolution of the bar strength parameter (see top of Figure \ref{fig:aa}). Radial dispersion velocities increase after the onset of the bar instability and there is a dip during the buckling instability. For non rotating and low spin halos the bars inside recover from buckling and continue to increase radial dispersion velocities, while the P90 disk is stagnant after buckling. The radial dispersion velocity does not reduce to its initial value in the P90 model after the buckling instability which explains why this disk does not under go a second bar instability---the disk is too hot.

The middle of Figure \ref{fig:disp} shows the time evolution of the vertical dispersion velocities normalized by $\sigma_z(t=0)$. For the P00 and P45 models, $\sigma_z$ is relatively constant until the onset of the buckling instability where there is a swift, large increase. $\sigma_z$ for these models appears to saturate and reach some equilibrium after the buckling instability.   The steepness of the growth from $\sigma_z$ at $t=0$ to $\sigma_z$ post buckling decreases with $\lambda$. This change in slope completely washes out the buckling instability in this plot for the P90 model. Between $t \sim 1.8$ Gyr and $t \sim 4.4$ Gyrs the vertical dispersion velocities linearly increase until they reach the final equilibrium value for the P90 model. This is a dramatic difference when comparing to the P00 and P45 models which look more like step functions rather than a slow and steady increases over large times.

The bottom of Figure \ref{fig:disp} displays the time evolution of the ratio of $\sigma_z/\sigma_r$. When the bar develops and $\sigma_r$ increase we see a dip in the ratio of dispersion velocities. Associated with the buckling instability is a swift increase in this ratio as the vertical dispersion velocities increase. For the P00 and P45 models the ratio reaches a maximum and then begins to decrease again, hinting at a future second buckling. For the P90 model we see the dip in the ratio followed by a small increase before the final jump with the buckling instability. Due to $\sigma_z$ slowly and linearly increasing before the buckling instability in the P90 model, at the time of buckling $\sigma_z/\sigma_r$ is larger when compared to the ratio for P00 and P45 just before buckling.
Over longer scales for the P90 model we see no evolution in the dispersion velocity ratio --- no second buckling or bar instability will happen for this stagnant disk.

 \begin{figure}
\centerline{
 \includegraphics[width=0.5\textwidth,angle=0] {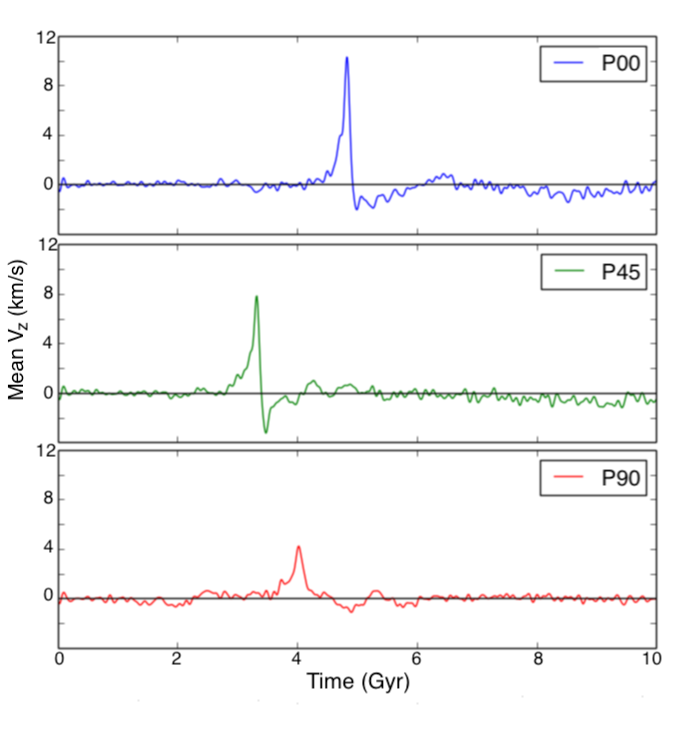}}
\caption{Time series of mean $v_z$ for all models. Outside the buckling instability the vertical velocities of the disk and bar should average to zero. The spike seen each model coincides the buckling instability.}
\label{fig:vz_time}
\end{figure}

Having looked at the variance in the velocities in the models we now turn our attention toward the evolution of the mean velocities plotted in Figure \ref{fig:vz_time}.  While the buckling wave is coherent we see a brief increase in mean value of $v_z$. Outside the buckling instability the mean velocity about the is zero. The spike of the maximum average $v_z$ coincides with the time of maximum buckling. We suggest that this measure acts as a better indicator of time of the actual buckling when compared to the bottom of Figure \ref{fig:aa} which is a measure of disk asymmetry. Note the trend seen in the height of the peaks for each model. Increasing halo spin limits the velocity reached by the buckled stars. This implies that the energy of the buckling is decreasing with increasing halo spin.

 \begin{figure*}
\centerline{
 \includegraphics[width=\textwidth] {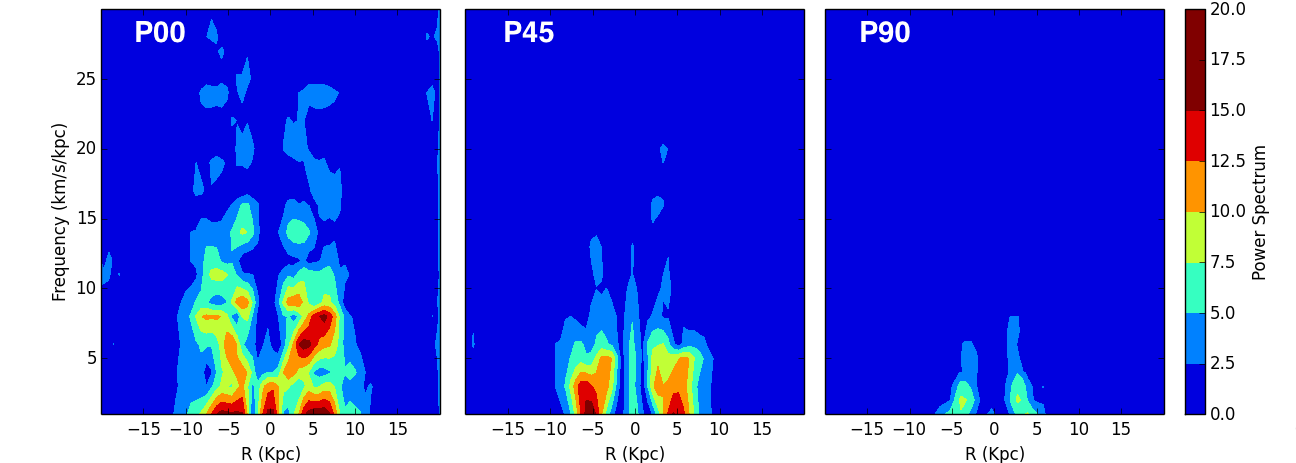}}
\caption{Fourier transform of mean $V_z$ distribution along radii $\pm 20$ kpc in the disk with data binned into $0.5$ kpc bins. This measurement was taken at all times relevant to buckling for each individual model. Color represents the buckling power at that position and frequency. More powerful buckling at higher frequencies is taken to be more energetic.}
\label{fig:freq_buckle}
\end{figure*}

To end this section, we compile a table of the resulting bar parameters from our simulations. $\Delta A2/A0$ and $t_b$ are calculated from the data in Figure \ref{fig:aa}. To measure $R_b$ and $z_b$ we calculate isodensity contours in the xy-plane and rz-plane respectively. The bar extends to the point were the ellipticity of the contour as decreased from its maximum by $15\%$. \citep[e.g.,][]{marti06, coll1}. The bar mass (stellar plus dark matter contribution) is measured in \citet{coll2}.

\begin{table}
\centering
\begin{tabular}{||c c c c||} 
 \hline
 Model & P00 & P45 & P90  \\ [0.5ex] \hline
$\Delta A_2/A_0$ & 0.19 & 0.29  & 0.37 \\ 
$R_{b}$ (kpc) & 9.4 & 6.2  & 6.1\\
$z_b$ (kpc) & 1.2 & 1.7 & 2.2 \\
$M_b$ $(10^{10}\,M_\odot)$ & 2.79 & 3.10 & 3.54 \\
$t_b$ (Gyr) & 0.2 & 0.2 & 0.7 \\[1ex] 
 \hline
\end{tabular}
\caption{Bar parameters for three models. $\Delta A_2/A_0$ is the drop in bar strength during buckling. $R_b$, $z_b$, and $M_b$ are the bar length, height, and mass at the time of buckling. Finally, $t_b$ is the  time required for the bar strength to go from the last pre-buckling maximum to the minimum bar strength.
}
\label{table:1}
\end{table}
\section{Discussion}
\label{sec:discussion}

We now apply the buckling instability analysis from Section \ref{sec:buck} to our models. For the development of the initial buckling perturbation we found the geometric and kinematic properties of the bar can effect the energy of the buckling. 

From the bottom Figure \ref{fig:disp} we note that the ratio of dispersion velocities for each model at the time of buckling is different, with the P90 model having the largest ratio of $\sigma_z/\sigma_r$.  We can relate the value of the buckling stress ($\tau_s/\tau_n$) to the ratio of vertical and radial dispersion velocities. The ratio of dispersion velocities measured at the time of buckling is $\sim 0.4$ for the P90 model, $\sim 0.32$ for the P45 model, and the ratio is $\sim 0.29$ for the P00 model. Looking only at the kinematic component we would expect the energy of the P00, and P45 models to be about equivalent with the energy being lower in the P90 model. However, the geometric properties should also effect the buckling energy. From Table \ref{table:1}, we calculate the slenderness of the simulated bars at the time of buckling and find that the most slender bar resides in the P00 model, followed by the P45 model and then the P90 model. From the dynamical argument in Section \ref{sec_buck}, the more slender bars should buckle with more energy. By analyzing the energy of the buckling in our models we can determine the relative importance of the geometric and kinematic properties of the bar to the buckling energy.

To quantify the energy of the buckling wave in each simulation, we start by finding the average $v_z$ in $0.5$ kpc bins for every relevant time step during buckling for each model. Next, we apply a Fourier transform to this data and plot the color map of the power spectrum which is shown in Figure \ref{fig:freq_buckle}. The frequency of the buckling wave is plotted for all relevant radii. Color denotes the power of the buckling wave on scale from blue (least powerful) to red (most powerful). Energetic bucklings will be more powerful at more radii than a less energetic buckling. If slenderness were the only predictor of buckling energy we would expect the bucklings of the P45 model, with a slenderness ratio of 3.6, to be on a similar scale to the P90 model, with a slenderness ratio of 2.8. However, we see that the bucklng is much less powerful in the P90 model compared to the P45 model. These models with similar geometric properties but very different kinematic profiles allow us to see that the kinematic properties are much more important to the initial buckling profile. We find the most powerful buckling in the P00 model, showing deeper reds and larger extent than the other two models. Note, the bar in the P00 model is longer than the other two models. As noted above the entire length of the bar is involved in the buckling perturbation which is why this model is spread along more radii than other models. From this figure, we find the kinematic properties of the bar are the most important for determining the buckling energy. The smaller the ratio of $\sigma_z/\sigma_r$ at the moment of buckling the stronger the pressure of the buckling force.

We now turn to the second phase of buckling and study the Landau damping in the simulated bars. We note that we do not expect these simulations (or the mixing within them) to be dependent on the particle number, $N$. Although too few particles in the disk will allow the bar to decay after buckling due to phase mixing effects we do not expect that to be relevant here. Particle number was explored in \citet{dub09}. The particle minimum needed to avoid those saturation effects is most readily seen in Figure 17 of their paper. We greatly exceed this minimum in our simulations.

 \begin{figure}
\centerline{
 \includegraphics[width=0.5\textwidth] {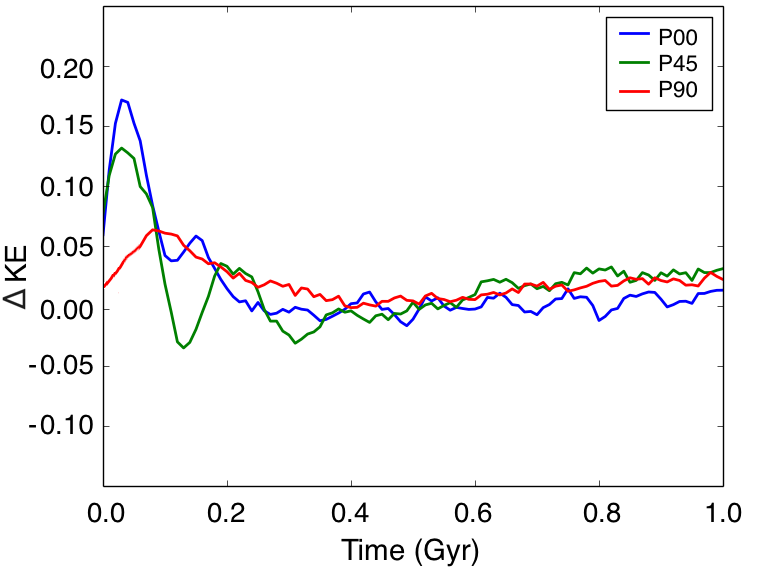}}
\caption{Calculated change in kinetic energy (KE) in the bar of each model normalized by the pre-buckling KE. The buckling instability happens at different times for each model. $t=0$ in this plot is the initial buckling time for easier comparison between models.}
\label{fig:data_mod}
\end{figure}

In Figure \ref{fig:data_mod} we compare the change in kinetic energy in each simulated bar from the time step just before buckling to long after the buckling has subsided.  The buckling time of each model is set to $t=0$ in this plot for better comparison between the different models. The frequency and amplitude of the wave is higher in the P00 model (blue) and P45 model (green) when compared to the P90 model (red). The larger the initial energy of the perturbation the more quickly it is damped out. The P00 and P45 models see more oscillations while the P90 model takes much more time to complete a single oscillation.

Why should the timescale of buckling effect bar evolution? A second process develops at the edge of the bar near corotation while the buckling instability is in progress. It has been known for some time that strong bars generate chaos \citep{cont81}  and the development of chaos during the buckling instability has been well documented in \citet{marti04}. The authors found the weakening of the bar happens at large radii and then propagates inward due to the increase in chaos during buckling. The strength of the bar sharply decreases due to the process which is happening simultaneously to the Landau damping. If buckling is lengthened more chaos will be introduced to the bar which will weaken the bar further when compared to bars that had a shorter buckling phase. More chaos induced in the outer region of the bar makes the radial bar orbits less coherent further decreasing the bar strength as seen in Figure \ref{fig:aa}.  The important result here is that a more volatile buckling instability found in the P00 model, and to a lesser extent in the P45 model, is not given enough time to dissolve the bar. In fact, the bar quickly recovers and continues to grow. The slow and gentle buckling found in the P90 model does, however, result in a dissolved galactic bar.

\section{Conclusions}
\label{sec:conclusion}
We have analyzed the buckling instability of a self-gravitating bar followed by the Landau damping that washes out the perturbation. We compare these results to three simulations with different kinematic and geometric properties in an attempt to discover what effect these properties have on the buckling instability and resulting disk evolution. Our results are summarized in the following paragraphs.

We found that the slenderness ratio of the galactic bar effects the energy of the buckling instability, with more slender bars having a more energetic buckling. For these models we considered the trapped dark matter component to contribute to the thickness of the galactic bar but other factors could play a roll in changing the bar parameters before buckling. For example, the thickness of the stellar disk can delay the onset of the bar instability and change the results of the buckling instability (e.g. \citet{aum17}).

Of more importance to the energy of the buckling perturbation, we found that there is is no simple correlation between a certain value of $\sigma_z/\sigma_r$ and the buckling instability. As seen in the bottom of Figure \ref{fig:disp} the time a stable barred disk can live below the ratio of $\sigma_z/\sigma_r < 0.4$ depends on many bar properties. We do find that a larger ratio of dispersion velocities will inhibit the energy of the buckling instability and plays a much larger roll when compared to the geometric properties.

Finally, an increase in the buckling time scale leads to a larger bar strength loss  ($\Delta A_2/A_0$) during the buckling instability. A weaker buckling energy will increase the time required for Landau damping to remove the perturbation. This allows the chaos that develops at the end of the bar to move to smaller radii, further dissociating bar orbits.  In our models, the bar parameters were changed by the initial conditions of the halo. There is evidently some turnover point for these models between $\lambda=0.045-0.09$ where the buckling timescale is lengthened and a large enough fraction of bar orbits become dissociated preventing the bar from recovering from buckling.  The work of \citet{coll1} allows us to limit the value of $\lambda$ for these models to $\lambda=0.045-0.06$ However, changing halo spin is not the only way to effect the initial conditions of the buckling instability.

We have shown that a nonviolent buckling can dissolve a strong bar. This result provides a possible answer to the question in the introduction. How can some disk galaxies avoid the bar instability for Hubble time? Perhaps, they formed a bar and following a weak buckling the disk was too hot to undergo a second bar instability. This idea is supported by observations of unbarred disks which are found to be kinematically hotter than their barred counterparts \citep{sheth12}. 

\begin{figure*}
\centerline{
 \includegraphics[width=\textwidth] {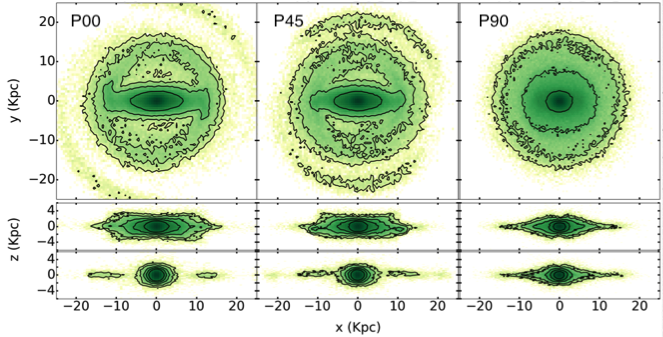}}
\caption{Linear isodensity contours of a 1 kpc slice of the stellar component for each simulation. Top row is face-on, middle is a view of the major axis and the bottom row is the bar viewed from the minor axis.
$\lambda$ increases to the right. We see a reduction in bar strength, disk radii, and spiral arm activity with the increase of $\lambda$ in the face on view. Edge on the B/P bulge shape decreases with the P90 models major and minor axis being nearly indistinguishable.}
\label{fig:cont}
\end{figure*}

While a weak buckling results in bar dissolution a violent buckling produces an edge-on B/P bulge. The first observational measurements of B/P bulges with redshift have been published by \citet{kruk19}. The authors found the fraction of bars with B/P bulges increased from $10\%$ at $z=1$ to $70\%$ at $z=0$. They find that buckling is suppressed in thick disks, not because of disk thickness but because of higher vertical dispersion velocities. This finding agrees with our results; in models with larger vertical dispersion velocities at the time of buckling we find a less obvious B/P bulge. We end by showing the observational results of differing buckling instabilities. In Figure \ref{fig:cont} we plot isodensity contours of the stellar disk surface densities for the face-on and edge-on disks. Bars that undergo a violent buckling host a B/P shaped bulge, while those that have an elongated weak buckling host a dissolved bar and no B/P bulge edge-on.

This work demonstrates the diverse stellar bar evolution that can follow buckling of varying initial conditions. However, we have not developed a comprehensive tool to predict the buckling instability parameters or the time of the initial symmetry breaking. Additional theoretical work must be done in order to gain a complete understanding of the buckling instability.

\section*{Acknowledgements}
We gratefully acknowledge support from NASA grant 80NSSC17K0720. This manuscript was greatly improved from suggestions made by Isaac Shlosman, Ann-Marie Madigan, and the reviewer Daniel Pfenniger. Thanks to Jeremiah Hudson for helpful mechanics discussions. This work utilized the RMACC Summit supercomputer, which is supported by the National Science Foundation (awards ACI1532235 and ACI-1532236), the University of Colorado Boulder, and Colorado State University. The Summit supercomputer is a joint effort of the University of Colorado Boulder and Colorado State University.

%\bibliographystyle{mn}
%\bibliography{MyRef}

\appendix

%%%%%%%%%%%%%%%%%%%%%%%%%%%%%%%%%%%%%%%%%%%%%%%%%%

\end{document}